\documentclass[%
 reprint,
 amsmath,amssymb,
 aps,
]{revtex4-2}

\usepackage{amsmath}
\usepackage{graphicx}
\usepackage{dcolumn}
\usepackage{bm}
\usepackage{lipsum}
\usepackage{booktabs}
\usepackage{caption} 
\usepackage{subfig}
\usepackage{graphicx}

\begin{document}

\preprint{APS/123-QED}

\title{Beamline Steering Using Deep Learning Models}

\author{Dexter Allen}
 \affiliation{San Jose State University, San Jose, CA, USA.}
\author{Isaac Kante}%
\affiliation{Long Island University, NY, USA.}

\author{Dorian Bohler}
\affiliation{
SLAC National Accelerator Laboratory, Menlo Park, CA, USA.
}%

\date{\today}

\begin{abstract}
Beam steering involves the calibration of the angle and position at which a particle accelerator's electron beam is incident upon the x-ray target with respect to the rotation axis of the collimator. Beam Steering is an essential task for light sources. The Linac To Undulator is very difficult to steer and aim due to the changes of each use of the accelerator there must be re-calibration of magnets. However with each use of the Beamline its current method of steering runs into issues when faced with calibrating angles and positions. Human operators spend a substantial amount of time and resources on the task. 
We developed multiple different feed-forward-neural networks with varying hyper-parameters, inputs, and outputs, seeking to compare their performance. Specifically, our smaller models with 33 inputs and 13 outputs outperformed the larger models with 73 inputs and 50 outputs. We propose the following explanations for this lack of performance in larger models. First, a lack of training time and computational power limited the ability of our models to mature. Given more time, our models would outperform SVD. Second, when the input size of the model increases the noise increases as well. In this case more inputs corresponded to a greater length upon the LINAC accelerator. Less specific and larger models that seek to make more predictions will inherently perform worse than SVD.
\end{abstract}

\maketitle


\section{Introduction, Overview \& Motivation}


Light sources refer to types of particle accelerators that generate strong beams of X-ray, ultra-violet or infrared light. These enable scientists to probe inside matter at nano- and even micro-scales, and at ultrafast speeds. Light-producing particle accelerators function differently from particle accelerators employed to investigate fundamental particles. Light sources employ the particle beam to emit light directly, as opposed to colliding particles and seeing what emerges. Through the use of an apparatus known as an undulator, that generates an alternating magnetic field, the beam path experiences a sequence of brief oscillations. The particles release photons whenever the route is bent. An undulator may include several of these oscillations, and the combined photons released from all of them create a powerful beam that is around one billion times brighter than a normal medical X-ray equipment. Numerous undulators may be accommodated and hundreds of different experiments can be served simultaneously by each light source. Certain electromagnetic spectrum ranges are used by light sources to function. All of the many forms of magnetic and electrical energy that exist in the cosmos are included in this spectrum. Depending on the magnitude of the waves they travel in, they are separated. A very little portion of this spectrum is used by humans to view visible light. Other portions of this spectrum include infrared, ultraviolet, and X-ray radiation, each having a distinct range of wavelengths. There are several uses for light at different wavelengths. Generally speaking, atomic structure may be seen at wavelengths shorter than visible light, such as the short end of X-rays. This implies that X-rays are capable of identifying the constituents of a material sample. The lowest wavelength X-rays, referred to as hard X-rays, are perfect for locating individual atoms inside of molecules or crystals. UV light and longer-wavelength X-rays, referred to as soft X-rays, are useful tools for researching chemical processes. The study of atomic vibrations in molecules and materials can benefit from the use of infrared light. Terahertz waves, the longest infrared light, are helpful for studying specific kinds of electronic structure, or the arrangement of electrons and their energy around an atomic nucleus. Light sources affect practically every field in study. They make it possible for scientists to find novel materials for quantum materials, microelectronics, solar panels, and batteries. To learn how to create better materials, they might examine how a material is formed or deteriorates over time. They have access to information that is not normally available and can look inside functioning equipment, even at nanoscales. They can capture 3D images of cells and other biological systems, providing insight into the fundamental functions of life. They are even able to determine the structure of extremely complex compounds, such as proteins. The LCLS (Linac Coherent Light Source) facilitates thousands of experiments every year aiding understanding of phenomena such as photosynthesis~\citep{photosystem}, molecular interactions toward drug discovery~\citep{drug_discover}, etc.


Beam steering \citep{metcalfe2007physics, gao2018quantification} is the process of calibrating the angle and location of the electron beam incident upon the x-ray target by a particle accelerator with regard to the collimator's rotating axis. Reliability of the dosage profile's form depends critically on precise beam steering.  Particle accelerator systems need beamline steering. Beamline steering is essential to these systems because it guarantees that the beam stays on its planned paths and target, which is crucial for the outcomes of operations and experiments. Beamline steering is crucial for a number of reasons, including safety, error correction, limiting beam loss, preserving beam stability, and minimizing interactions among accelerator components. The success of several applications in a variety of sectors, the caliber of scientific research, and the safety of accelerator operations are all greatly impacted by beam steering. Beam steering is also essential for minimizing errors and flaws in the accelerator's magnetic and radio frequency systems, which enables peak performance and the effective completion of numerous tests and applications in disciplines like materials science, particle physics, and medical treatment.


In mathematical terms, the beam steering problems can be characterized via $N$ Beam Position Monitors (BPMs) \cite{forck2009beam, smith1997beam} providing the position of the beam at a certain location $i$, as $u_i$; and $M$ correctors providing dipole deflections $\theta_j$. These can be related via a response matrix, $R$, such that $\Delta u = R \Delta \theta$. Thus, given a set of measured current beam positions, $u_m$ we wish to find corrections $\theta_c$ such that the norm of $u_m + R \theta_c$ is minimized. Currently, the most common approach for this steering problem is the use of Singular Value Decomposition (SVD), where we use the transformation to create a basis in corrector space such that the action of the response matrix maintains this orthogonality. While this is an effective technique, it has some limitations. For instance, it ignores the non-linear effects of the quadropole fields that are important at high amplitudes. Thus, the SVD approach is only applicable at low corrector strengths. 


Owing to the importance of non-linear effects especially at higher amplitudes we focus on using non-linear models for the task of beam steering. As deep learning models are universal approximators they can model arbitrarily complex, non-linear relationships. So deep learning models can handle complicated and dynamic real-time data and they presents a viable alternative to beamline steering using SVD approaches. This makes it possible to make exact modifications to preserve beam stability and improve trajectory. Large amounts of data may be processed by deep learning models, which can also identify patterns that linear approaches would be unable to. Real-time analysis of this data, identification of departures from the predicted path, and autonomous generation of optimal beamline component and magnet steering settings are all made possible by deep learning models. Their ability to adjust to shifting circumstances and react swiftly to disruptions improves beam quality and reduces the likelihood of beam loss. In this investigation we focus on developing deep learning models for beam steering.


Anomaly identification, system modeling, virtual diagnostics, data analysis, and beamline tuning are a few uses of machine learning in beamlines. Real-time machine learning models have the potential to detect anomalies or unforeseen actions in accelerator systems, including beam instability or equipment malfunctions \cite{fol2017evaluation}. These systems can start safety procedures or stop operations when anomalies are detected to protect personnel and equipment. By using machine learning to build models that mimic the behavior of accelerator systems, beam trajectories and performance can be predicted and the effects of various parameters and components can be comprehended \cite{gupta2021improving, mishra2021uncertainty}. These models support system design and optimization. Virtual instrumentation and diagnostics employ machine learning to create virtual representations of accelerator systems and components. Through the testing of various conditions without the requirement for actual tests, these simulations can provide insights into beamline behavior, aiding in system design and diagnostics \cite{pang2015advances}. Large-scale data generated by accelerator systems is subjected to advanced data analysis techniques and machine learning to derive meaningful insights. They are able to identify trends, correlations, and patterns in data to improve beamline guiding, find solutions to issues, and increase efficiency.

This paper is arranged as follows. In the first section we give an introduction to the task and challenges of beam steering and the manner in which deep learning may be beneficial there. In the second section we provide the mathematical details of the current method and for deep learning. We also cover the datasets used. In the third section we introduce the forward and inverse neural network models for beam steering and discuss their results. In the final section we give a summary of the analysis and results.

\section{Mathematical \& Computational Details}

\subsection{Overview of the SVD approach}
Singular Value Decomposition is an essential element in linear algebra and is mainly used to simplify matrix operations and reveal important characteristics of a matrix. The introduction of SVD began in the early 20th Century by Erhard schmidt in 1907. SVD was further developed in 1940s-1950s by mathematicians like G.E.P. box, David Hilbert, and Carl Eckart.\cite{eckart1936approximation} In the beginning of the 1960s and 1970s with the arrival of computers and numerical analysis, SVD became a key concept for solving computational problems.\cite{strang2016introduction} James H. Wilkinson and C.L.Lawson were the two pioneers in creating numerical algorithms for SVD to be more accesible for applications. Spanning into the 21st century SVD was utilized in data science, statistics, and machine learning. Methodologies like Principal Component Analysis (PCA) and Latent Semantic Analysis (LSA) utilize SVD to analyze and reduce data dimensions. The utilization of SVD has a significant impact on modern data-driven fields.\cite{bishop2006pattern}

SVD simplifies a matrix by splitting them into three other matrices. For example given a Matrix \(A\) of the size \(m\) \(x\) \(n\), SVD breaks \(A\) into three matrices\begin{equation}
    A = U\sum V^T 
\end{equation}
\(U\) comprises the left singular vector of \(A\). The vector are orthonormal meaning they are perpendicular and have unit length.  \(\sum\) contains the singular values of \(A\) on its diagonal, arranged in descending order providing insights into the matrix's rank and its inherent structure. \(V^T\) contains the right singular vectors of \(A\). Similarly to \(U\) the columns of \(V\) are orthonormal vectors.

For beamline steering, the Singular Value Decomposition (SVD) approach provides a potent means of rerouting particle paths in accelerators. The objective of this method is to reduce, in a least-squares sense, the difference between the target orbit and the measured beam orbit. The system makes use of several BPMs (Beam Position Monitors) and correctors. BPMs are a flexible, non-destructive device for the determination of the beam’s centre of mass. The correctors (or corrector magnets) induce orbit shifts in the beam by changing the magnet's settings and currents. The orbital key equation

The SVD adjustment is:

\[\theta = (R^T R)^{-1} R^T \Delta x\]

Here, \(R\) is the orbit response matrix, \(\theta\) is the vector of desired changes to the corrector magnets, and \(\Delta x\) is the difference between the target orbit and the measured orbit. The Singular Value Decomposition (SVD) of the orbit response matrix, as discussed in "Beam-based correction and optimization for accelerators" (Huang, 2020), is expressed as:

\[
R = U \Sigma V^T
\]

where \(\Sigma\) is a diagonal matrix of singular values and \(U\) and \(V\) are orthogonal matrices. The singular values show how well various orbital modification methods work.

Either an under-constrained or an over-constrained system can have a unique least-squares solution thanks to the SVD approach. The pseudo-inverse of \(R\) is taken into consideration to further improve the solution, particularly in cases where the number of correctors exceeds the number of BPMs (Beam Position Monitors). Singular values are used to calculate the pseudo-inverse, which guarantees a stable result.

BPMs can be weighted differently in the orbit control system according to their relevance thanks to the SVD technique. This adaptability is useful when precise control is required at particular points along the beamline. Empirical examples of the method's application include orbit error correction in accelerators like SPEAR3. The results demonstrate that orbit distortion can be significantly reduced with a very modest number of single value modes. The method is further expanded to incorporate BPM weighting parameters, enabling orbit control to be fine-tuned according to particular needs at various sites.

\subsection{Overview of Deep Learning}
Within the ecosystem of artificial intelligence (AI) and machine learning (ML), deep learning (DL) stands out as one of the foundational and core technology of today. DL is  a subset of ML and AI, and is viewed as an AI function that mimics the human brain. The difference in ML and DL being that ML models often need manual feature extraction, while DL models automatically learn features through multiple layers of representation.\cite{zhao2023deep} 

DL has the same workflow properties as ML when building a model. The three processing steps consist of data understanding and preprocessing, DL model building and training, and validation and interpretation. DL unlike ML modeling, feature extraction is automated instead of manual. ML is not dependent on a large amount of data to build a data-driven model like DL. DL typically performs poorly on small data volume. DL also requires huge computational operations while training a model with large data volume. A Graphical Processing Unit (GPU) is necessary over a Computer Processing Unit (CPU). Feature engineering is the process of extracting features from raw data. Unlike ML, DL decreases the time and effort required to construct a feature extractor for each problem.\cite{zhao2023deep} DL is very time consuming and computational resource heavy. Since DL trains on large datasets and volume with large number of parameters the training process can take more than one week to a month. On the contrary DL testing takes little time to run when compared to ML models. DL interpretability is an important factor and can be difficult to explain how results were determined. This is referred to as "Black-Box". ML algorithms have rule based techniques (IF-THEN) for making decisions that are easy to understand results and findings.\cite{zhao2023deep} 

Overall DL performance grows exponentially with large data volume. DL excels with large amount of data and its capacity to process vass amounts of features to build data driven models. Many libraries and sources like PyTorch and TensorFlow utilize matrix and tensor operations as well as computing gradients and optimization. These libraries give the necessary function for implementing and building a DL model.

\subsection{Overview of Data used}

\subsubsection{Archival Data}
This project contained two different types of data in order to train the models. The neural network developed for the inverse mapping problem used both archival and simulation data while the neural network developed for the forward mapping problem was only trained on archival data. The archival data was acquired from EPICS (Experimental Physics and Industrial Control System) archiver \cite{dalesio1991epics}. EPICS is a software designed to control and monitor complex scientific experiments and industrial processes. EPICS was originally developed by SLAC (Stanford Linear Accelerator Center) and is used widely by many others in the field of sciences.\cite{epics} The archival dataset is made up of real-world, historical data that captures genuine system activity. Epics has allowed us to retrieve specific timestamps from the data as well as the format of the data. Collecting the archival data required calling the method lcls archiver which supplies the parameters PVLIST and EPICS timestamp. 
\subsubsection{Simulation Data}
The simulation data was obtained by a python library called PyTao (Python Tools for Accelerator Optics) created to design simulations and analysis of particle accelerators.\cite{pytao} Simulation data has played a critical role in this project as it has shed light on the inverse models weakness. Although the archival data was made up of real system activity the simulation data generated different scenarios from a wide range. Although simulation data does not reflect real world details, it gives a wide diversity. 

\subsubsection{Inputs and Outputs}
In this project we test two different varieties of mapping. The forward mapping is named as it follows the paradigm of the SVD application. This mapping tries to predict the X- and Y-corrector settings based on the BPM measurements and the quad settings. The inverse mapping predicts the BPM measurements based on the corrector settings and the quad settings. 

During this project the features for both the forward and the inverse model are the same except for the inputs and outputs. The forward model was built on 46 features which included the following inputs and outputs, 33 inputs being  XBPMS (X Beam Position Moniters) YBPMS ( Y  Beam Position Moniters) and QUADS (Quadruple Magnets) \cite{lazzaro2008field, dayton1954measurement}, and 13 outputs being XCOR (X Corrector) and YCOR (Y Corrector). The inverse model was built on 123 features which included 73 inputs being XCOR, YCOR, and QUADS. The outputs were 50 XBPMS and YBPMS.

\subsection{Details about the hyperparameters}
It is well known that hyperparameters play an important role in deep learning model building and structure \cite{pon2021hyperparameter}. During this project the two models had distinct hyperparamaters. For the neural network developed for the forward mapping problem model the hyperparameters chosen were an 8 layer Feed Forward Neural Network and ReLU activation, hidden size of 64 nodes, learning rate 0.0001 and an epoch of 300. For the neural network developed for the inverse mapping problem the hyperparameters chosen were 7 layer Feed Forward neural network and ReLU activation, hidden size of 512, learning rate of 0.001, Adam (Adaptive Moment Estimation) optimization \cite{kingma2014adam}, and an Epoch of 300.  
\begin{figure*}
    \centering
    \includegraphics[width= 0.8 \linewidth]
    {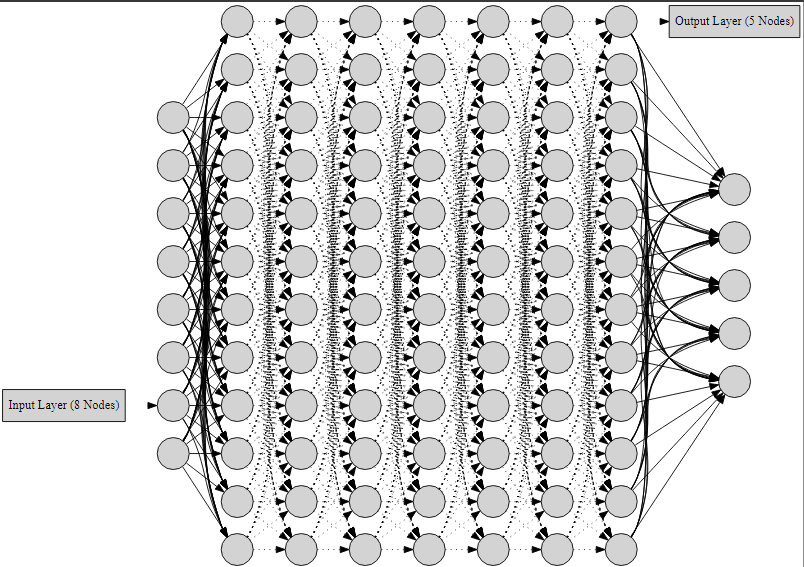}
    \caption{The neural network model developed for the inverse modeling paradigm:  Neural Network Model Architecture}
    \label{fig rmse}
\end{figure*}

\section{Results \& Discussion}

\subsection{Forward Neural Network Modeling Paradigm}

The forward model predicted the X- and y-corrector settings based on the beam position monitor readings and the quadropole magnet settings. This Feed Forward Neural Network has 8 hidden layers with 64 Nodes each. The model was trained on 75 million data points. To train the model, we looked at the top $25\%$of runs from March 2021 to November 2022. This analysis resulted in 200 million data point samples. We then gathered relevant data at the LTU during these times, giving us 75 million data points to train the model with.

For the forward mapping, the current methods RMSE (Root Mean
Squared Error) can range from 0.016 to 0.008. The deep learning model developed for the forward mapping task has an RMSE of 0.0034, which is consistently better than the current state of the art. 

\begin{figure*}
    \centering
    \includegraphics[width= 0.5 \linewidth]
    {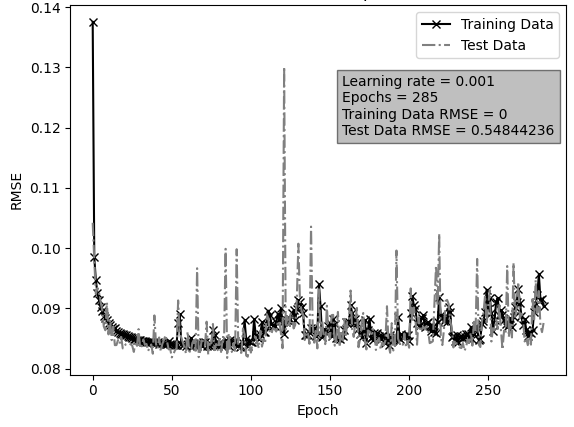}
    \caption{Training history for the neural network model developed for the inverse modeling paradigm: Root Mean Squared Error For Test and Training data}
    \label{fig rmse}
\end{figure*}

\begin{figure*}
    \centering
    \subfloat[]{
    \includegraphics[width= 0.45\linewidth]
    {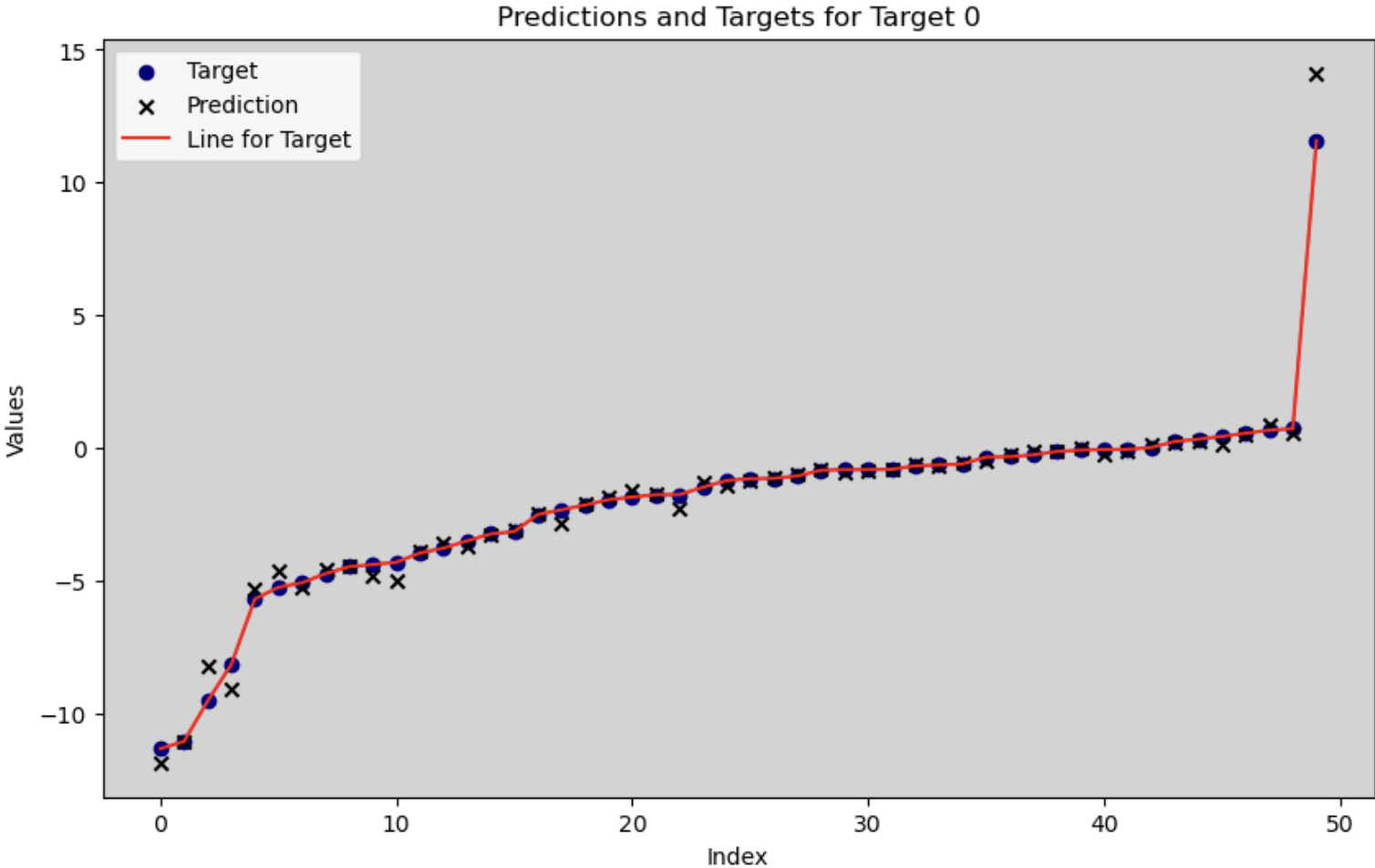}
    \label{fig inverse model pred arch}
    }
    \hspace{0.08\linewidth}
    \subfloat[]{
        \includegraphics[width=0.45\linewidth]
        {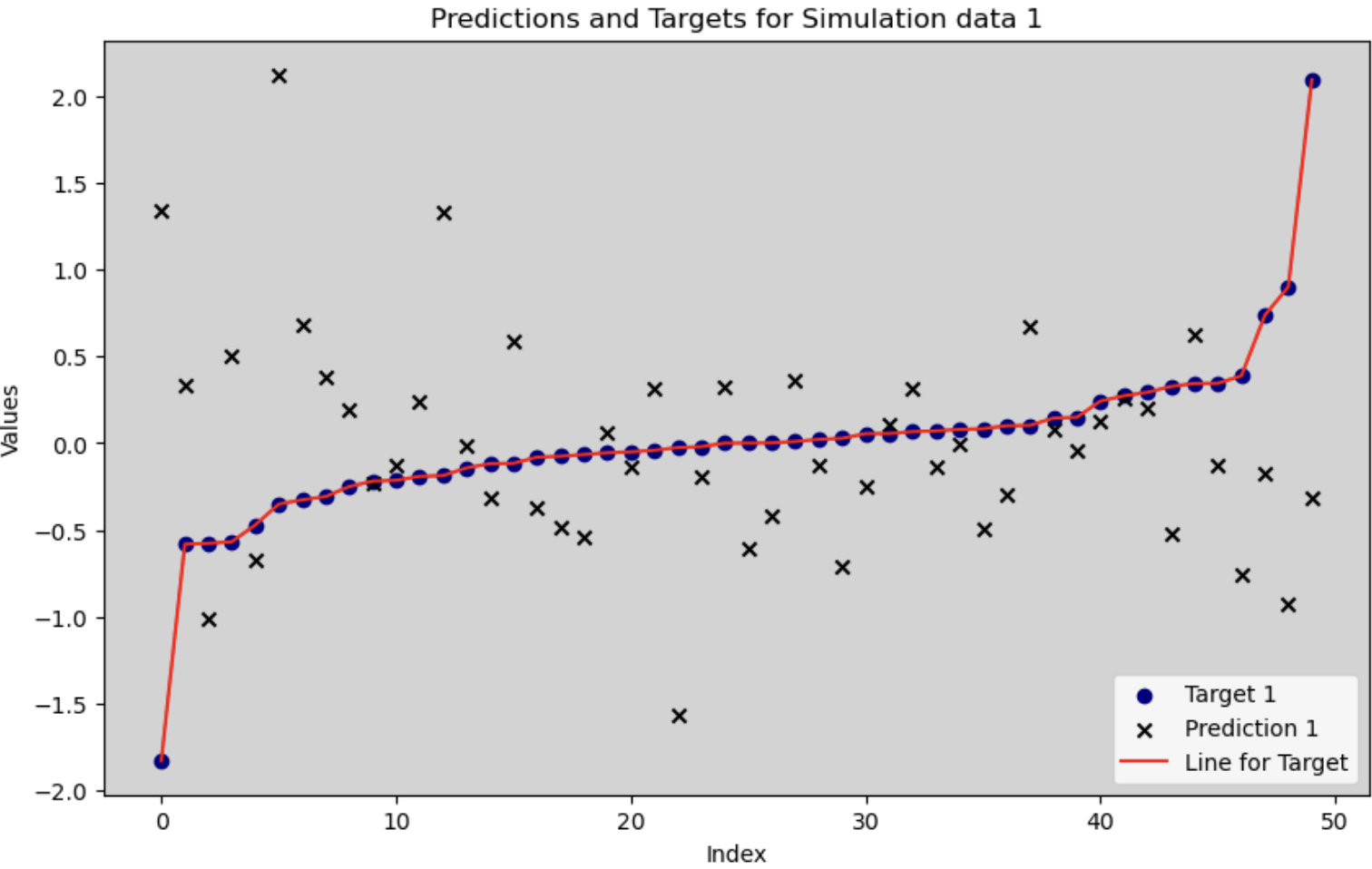}
        \label{fig sim 1}
    }
    \hspace{0.08\linewidth}
    \subfloat[]{
        \includegraphics[width=0.45\linewidth]
        {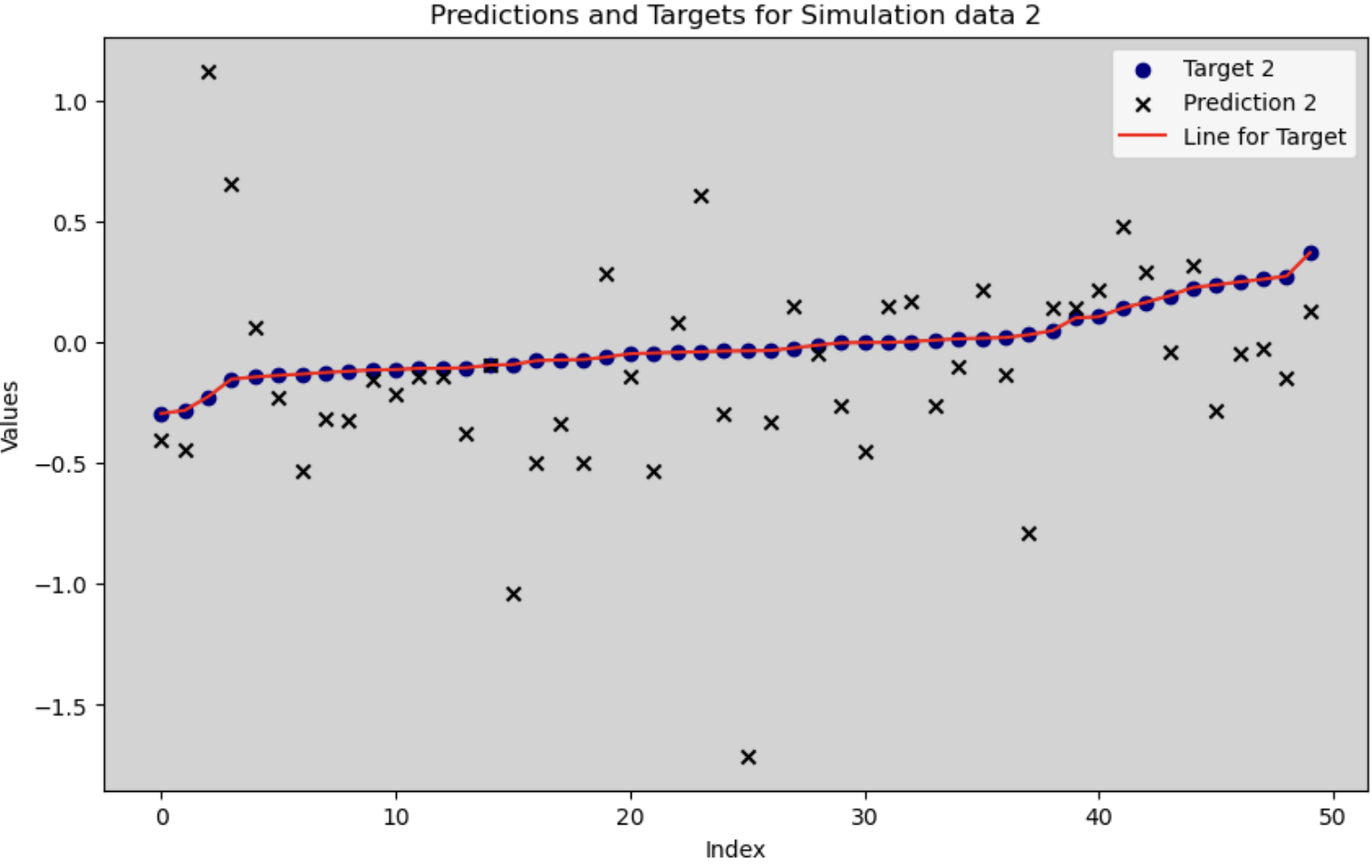}
        \label{fig sim 2}
    }
    \hspace{0.08\linewidth}
    \subfloat[]{
        \includegraphics[width=0.45\linewidth]
        {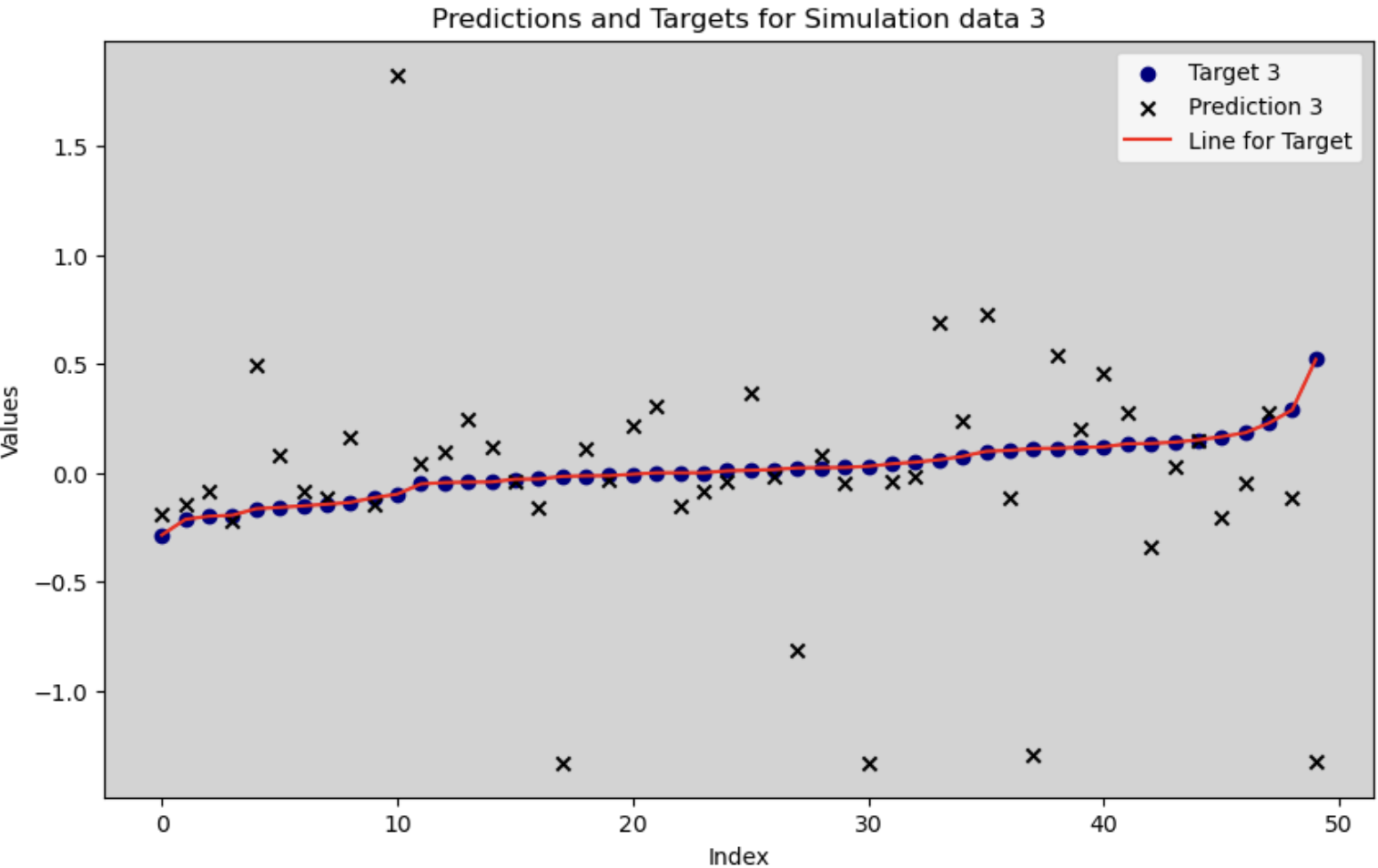}
        \label{fig sim 3}
    }
    \caption{Inverse model predictions on archival data and simulation data}
    \label{fig:combined}
\end{figure*}

\subsection{Inverse Neural Network Modeling Paradigm} 
The inverse deep learning model predicted the Beam Position Monitor readings based on the quadropole magnet settings and the x- and y-corrector settings. The Feed Forward Neural Network (FFNN) was created using Pytorch machine learning library.\cite{pytorch} This model did training on a larger set of archival data than the neural network developed for the forward mapping problem model with the total number of data points being 1697935 rows and 123 columns. The model was comprised of four components, 7 linear layers, input size, hidden size and output size. The hidden size selected for this model was 512 neurons that was used for applying rectified linear unit (ReLU) activations after each hidden layer to introduce non-linearity. Two functions called \(training step\) and \(validation step\) were created to monitor the training and validation rmse loss for the models predictions and targets for a given batch. The models hyperparemeters included a learning rate (lr) of 0.001, 300 Epoch and Adaptive Moment Estimation (Adam) optimizer to update the models parameters using the specified lr. As the training for the model went on for each batch in the training dataset, gradients were zeroed, backpropagation is carryed out and the optimizer updates the models weights. After training the model has not shown any clear sign of accuracy or understanding as it stayed halted at a training loss of 0.007009 and  Validation loss of 0.006707. The results of this model were calculated by three metrics Mean absolute error(MAE), Mean squared error(MSE) and Root Mean Squared Error (RMSE). The results for all three metrics were 0.159 MAE, 0.308 MSE and 0.555 RMSE. These metrics indicate that there is some room for improvement in the model and there are high errors. This metrics for the simulation data on this model were worse than the archival. This also is due to the models inability to interpret the complexity and pattern of new data as shown in Table \ref{tab:model_performance}. This shows that the neural network developed for the inverse mapping problem is not able to generalize its performance. It has much better performance for the data that it is trained on (archival data) than for new data that it has not trained on (simulation data) as presented in Figure \ref{fig:combined}. But this can also be because of the covariate shift \cite{nair2019covariate, sugiyama2012machine} between the real world archival data and the simplified simulation datasets.

\begin{table}[h] 
    \centering
    \caption{Inverse mapping neural network Model Performance Metrics on Archival Data and Data from Simulations.}
    \begin{tabular}{lccc} 
    \toprule 
    Data & MAE & MSE & RMSE \\ 
    \midrule 
    Archival Data     & 0.159 & 0.308 & 0.555 \\ 
    Simulation Data 1 & 0.606 & 0.816 & 0.903 \\
    Simulation Data 2 & 0.311 & 0.200 & 0.448 \\
    Simulation Data 3 & 0.340 & 0.324 & 0.569 \\
    \bottomrule 
    \end{tabular}
    \label{tab:model_performance}
\end{table}

\subsection{Comparison of the Forward and the Inverse Mapping Paradigms}
We see that compared to the inverse neural network, the forward neural network model performs significantly better. Even after a thorough search for an architecture and fine-tuning of hyperparameters, this still occurred. We think that the inverse mapping's ill-posedness is the source of this. Given the BPMs and quad measurements, predicting the corrector settings is a well-posed issue with a special solution. Inadequate model accuracy results from the inverse mapping, which predicts the BPM measurements given the corrector and quad parameters. It is not consistently well conceived.

\section{Summary \& Conclusions}
The task of beam steering is crucial for the scientific throughput of light sources. Beamline steering is now done by skilled operators. Light sources would benefit from faster, more dependable, and automated beamline steering to increase the number of tests that can be carried out at the facility and the caliber of the outcomes. We have a lot of data to learn this mapping, therefore machine learning models make sense for our application. The best option is to use deep learning-based models because of their accuracy and adaptability. Deep learning algorithms are not impervious to covariate movements and might occasionally provide responses that are overconfident.

In this study we analyze two different paradigms of using deep learning to guide the task of beam steering. The forward model predicted the X- and y-corrector settings based on the beam position monitor readings and the quadropole magnet settings. The inverse model predicted the Beam Position Monitor readings based on the quadropole magnet settings and the x- and y-corrector settings. The models provided extremely accurate responses when trained on the archive data. However, their accuracy was incorrect by at least a factor of two on simulated datasets. This demonstrates how deep learning models are vulnerable to changes in covariates. The deep learning models performed nearly ten times worse in all the experiments than the Singular Value Decomposition method currently in use for beamline guiding. 

We observe that the forward neural network model has far superior performance than the inverse neural network. This was even after exhaustive architecture search and hyperparameter tuning. We believe that this is because of the ill posed character of the inverse mapping. Predicting the corrector settings given the BPMs and quad measurements is a well posed problem with a unique solution. The inverse mapping of predicting the BPM measurements given the corrector settings and the quad settings is not uniformly well posed and causes inferior model accuracy.

\bibliography{apssamp}

\end{document}